\documentclass[10pt,final,twocolumn]{IEEEtran} 

\newcommand{\imgwidth}{1} 

\usepackage{ifthen}
\usepackage{graphicx}
\usepackage{amsmath}
\usepackage{amssymb}
\usepackage{setspace}
\usepackage{cite}

\usepackage{lastpage} 
\usepackage{eso-pic}  

\usepackage{lastpage} 
\usepackage{eso-pic}  
\usepackage{hyperref}
\AddToShipoutPicture{
\put(0,15){ 
\begin{minipage}{\paperwidth}
\footnotesize
\hspace{1in}\hspace{\oddsidemargin}
\hfill
\href{http://ieeexplore.ieee.org/xpl/articleDetails.jsp?arnumber=6882216}{\textbf{cite}}  \hfill
\thepage\,/\,\pageref*{LastPage}
\hspace{1in}\hspace{\oddsidemargin}
\vspace{10ex}
\end{minipage}
}}

\newcommand{\bb}[1]{\mathbf{#1}}
\newcommand{\mc}[1]{\mathcal{#1}}
\newcommand{\mE}[1]{\mathrm{E}\left[#1\right]}

\title{Theoretical Framework for the Optimization of Microphone Array Configuration for Humanoid Robot Audition}
\author{V.~Tourbabin*,~\IEEEmembership{Student~Member,~IEEE} and B.~Rafaely,~\IEEEmembership{Senior~Member,~IEEE} 
\date{}
\thanks{

The authors are with the Department of Electrical and Computer Engineering, Ben-Gurion University of the Negev, Be'er-Sheva 84105, Israel (email: \{tourbabv,br\}@ee.bgu.ac.il)}}

\begin{document}
	\maketitle
	\begin{abstract} 
An important aspect of a humanoid robot is audition. Previous work has presented robot systems capable of sound localization and source segregation based on microphone arrays with various configurations. However, no theoretical framework for the design of these arrays has been presented. In the current paper, a design framework is proposed based on a novel array quality measure. The measure is based on the effective rank of a matrix composed of the generalized head related transfer functions (GHRTFs) that account for microphone positions other than the ears. The measure is shown to be theoretically related to standard array performance measures such as beamforming robustness and DOA estimation accuracy. Then, the measure is applied to produce sample designs of microphone arrays. Their performance is investigated numerically, verifying the advantages of array design based on the proposed theoretical framework.  
\end{abstract}



	\section{Introduction} 
\IEEEPARstart{H}{umanoid} robotics appears to be a fast evolving field in recent years. Humanoid robots are studied and developed for various applications including service, welfare and entertainment. For a recent review of humanoid robots see, for example,  \cite{Akhtaruzzaman2010}. An important component of a humanoid robot, facilitating communication with the surroundings, is the auditory system. This system should be capable of performing various tasks related to audio communication and orientation, including sound localization \cite{Middlebrooks1991}, source separation \cite{Litovsky2012} and scene analysis \cite{Robart2005}.

The performance of a robotic auditory system depends on the complexity of the surrounding sound field, the way in which the surrounding sound field is acquired, and the algorithms operating on the acquired data. One of the most flexible and effective means for sound acquisition is based on microphone arrays; these have been successfully applied in robot audition tasks during the last decade. The applications include human-like robotic systems capable of sound source localization, tracking, and separation \cite{Strumillo2011,Keyrouz2006,Hornstein2006,Kim2013}, with the sound field acquired by only two microphones located at the ears. While these binaural systems are studied primarily because of their implementation simplicity and their human-like appearance, there is a growing interest in humanoid audition using more than two microphones. For example, the speech recognition system implemented on the HRP-2 robot \cite{Takahashi2010} uses a source separation algorithm that utilizes $8$ microphones distributed on the head of the robot. Another $8$-microphone array, installed in the Hearbo robot, was utilized for demonstrating an advanced source localization and separation algorithm \cite{Nakamura2013}. Another example is the source separation algorithm developed for the ROMEO robot that uses $16$ microphones distributed on a circle around the head \cite{Maazaoui2012}. All these applications can potentially benefit from an acoustically optimized array configuration. However, the literature concerned with the optimal design of array configuration for robots is limited. 

The performance of an auditory system that acquires the sound field by means of a microphone array depends, among other factors, on the array configuration, defined by the number of microphones and their positions.
From the microphone array processing literature it is generally known that performance improves with an increase in the number of microphones and may be significantly affected by their positions, e.g. overall array aperture, and distance between microphones \cite{VanTrees2002}. Furthermore, there are publications describing optimal microphone positioning to improve the side-lobe level \cite{Murino1996}, to increase the number of uncorrelated sources that can be distinguished \cite{Pearson1990}, and to improve the direction of arrival (DOA) estimation accuracy under various noise conditions \cite{Zhu2010,Tuladhar2011}. However, these methods are limited to linear arrays. A method for optimal sensor positioning for an arbitrary array geometry is described in \cite{Joshi2009}. But this is a narrow-band method and is only strictly applicable to waveform estimation when the source directions are known.
The most relevant study to the current work is reported in \cite{Skaf2011}, where a wide-band criterion is used for optimal positioning of two microphones on the opposite sides of a humanoid-robot head. However, the criterion that is used there is only  strictly appropriate to a pair of microphones; it is based on the extraction of features, such as interaural time difference (ITD) and interaural level difference (ILD), from the head related transfer functions (HRTFs) of the robot.

The approach adopted in this work is to first develop a wide-band measure of array quality for an arbitrary number of microphones and array geometry. Then, the measure is employed for array design and optimization. The measure proposed here reflects the amount of information regarding the surrounding sound field that is available to the array as a function of microphone positioning \cite{Tourbabin2013}. The amount of information is quantified based on the effective rank \cite{Roy2007} of the generalized head related transfer functions (GHRTFs) matrix that relates the sound field to the array measurements.

The paper is structured as follows. First, in sections \ref{sec:sII} and \ref{sec:sIII}, a measurement model is described, based on the GHRTFs that account for microphone positions other than the ears. Then, in sections IV and V, a measure of array quality is proposed \cite{Tourbabin2013} and its significance in beamforming and sound localization systems is investigated theoretically. Finally, in section \ref{sec:ex}, a method for obtaining the GHRTFs and examples of optimal microphone positioning are presented and discussed.

	\section{Background} 
\label{sec:sII}
A standard measurement model commonly used in the array processing literature is presented in this section as a background to the proposed method. Then, some formulations and notations related to beamforming and DOA estimation that are required in the following sections are outlined. 

\subsection{Measurement model} 
\label{sec:measmod}
A commonly-used model in the array processing literature relates the signal measured by the array microphones at a given frequency to the source signals that produce the measured sound field.
Consider a humanoid robot head submerged in a sound field produced by $D$ spatially separated sources. The complex sound pressure amplitudes at frequency $\omega_0$ measured by $L$ microphones positioned on the surface of the robot head can be written as:
\begin{equation}
	\bb{p}=\bb{As}+\bb{n},
	\label{eq:101}
\end{equation}
where 
\begin{equation}
	\bb{p}=[p_1(\omega_0)\,\,p_2(\omega_0)\,\,\cdots\,\,p_L(\omega_0)]^T\in\mathbb{C}^{L\times 1}
	\label{eq:101aa}
\end{equation}
holds the complex pressure amplitudes measured by the microphones and vector
\begin{equation}
	\bb{s}=[s(\omega_0,\Omega_1)\,\,s(\omega_0,\Omega_2)\,\,\cdots\,\,s(\omega_0,\Omega_D)]^T\in\mathbb{C}^{D\times 1}
	\label{eq:101ab}
\end{equation}
holds the complex signal amplitude of the sources, assumed to be positioned in the far field at a direction denoted by $\{\Omega_j\}_{j=1}^{D}$. Here $\Omega_j=(\theta_j,\phi_j)$ denotes azimuth $\theta_j$ and elevation $\phi_j$ in a spherical coordinate system \cite{VanTrees2002}, and $(\cdot)^T$ denotes the matrix transpose operator.
Matrix 
\begin{equation}
	\bb{A}=[\bb{a}(\omega_0,\Omega_1)\,\,\bb{a}(\omega_0,\Omega_2)\,\,\cdots\,\,\bb{a}(\omega_0,\Omega_D)]\in\mathbb{C}^{L\times D}
	\label{eq:101ac}
\end{equation}
is composed of the steering vectors of the array at frequency $\omega_0$, where steering vector 
\begin{equation}
	\bb{a}(\omega_0,\Omega_j)=[a_1(\omega_0,\Omega_j)\,\,a_2(\omega_0,\Omega_j)\,\,\cdots\,\,a_L(\omega_0,\Omega_j)]^T
	\in\mathbb{C}^{L\times 1}
	\label{eq:101ad}
\end{equation}
relates the signal of the $j^{th}$ source to the complex sound pressure amplitude measured by the $L$ microphones at frequency $\omega_0$. Vector $\bb{n}\in\mathbb{C}^{L\times 1}$ represents additive noise, whether present in the measured sound field or generated by the instrumentation.

\subsection{Beamforming}
One of the most common methods of microphone array processing is spatial filtering or beamforming.
Based on the measurement model presented above, the narrow-band response of a spatial filter to a far-field source signal of unit amplitude in direction $\Omega$ is described by its beampattern \cite{VanTrees2002}:
\begin{equation}
	B(\Omega)=\bb{w}^H\bb{a}(\Omega),
	\label{eq:101b}
\end{equation}
where $\bb{w}^H$ contains the beamformer weights and $\bb{a}(\Omega)$ is the array steering vector due to a single source. The dependence on frequency of $B(\Omega)$, $\bb{w}^H$ and $\bb{a}(\Omega)$ is omitted for notation simplicity. The operator $(\cdot)^H$ denotes the conjugate transpose.

One widely studied data-independent beamforming approach is the maximum-directivity beamformer \cite{VanTrees2002,Tourbabin2012b}. The beamformer weights for a given look direction $\Omega_l$ are given by:
\begin{equation}
	\bb{w}_{MD}^H=\frac{\bb{b}^H\bb{C}^{-1}}{\bb{b}^H\bb{C}^{-1}\bb{b}},
	\label{eq:101f}
\end{equation}
where $\bb{b}=\bb{a}(\Omega_l)$ is the steering vector at the array look direction and $\bb{C}$ is the following matrix:
\begin{equation}
	\bb{C}=\frac{1}{4\pi}\int\limits_{\Omega\in S^2}{\bb{a}(\Omega)\bb{a}^H(\Omega)}d\Omega.
	\label{eq:101d}
\end{equation}
The integral $\int_{\Omega\in S^2}{d\Omega}=\int_{0}^{2\pi}{\int_{0}^{\pi}{\sin\theta d\theta d\phi}}$ covers the entire surface of the unit sphere, denoted by $S^2$. This implies that $\bb{C}$ represents the average of $\bb{a}(\Omega)\bb{a}^H(\Omega)$ over all possible source arrival directions. Assuming that the average can be approximated by a finite summation and using the model in Eq. \eqref{eq:101}, leads to:
\begin{equation}
	\bb{C}\approx\frac{1}{D}\sum\limits_{j=1}^{D}{\bb{a}(\Omega_j)\bb{a}^H(\Omega_j)}=\frac{1}{D}\bb{AA}^H,
	\label{eq:101e}
\end{equation}
where $D$ is the number of directions used for the approximation.
Note that in order to obtain the maximum-directivity beamformer weights utilizing the approximation of $\bb{C}$ given in \eqref{eq:101e}, $\bb{AA}^H$ should be non-singular. Necessary conditions for this to be the case include placing the microphones at different positions and $D\geq L$. 

A widely accepted performance measure is that of beamformer robustness \cite{VanTrees2002}, representing the sensitivity of the beamformer to small perturbations in array weights and imprecise microphone positioning. Beamformer sensitivity is given by the squared Euclidean norm of the weight vector and for the maximum-directivity beamformer (see \eqref{eq:101f}) is given by:
\begin{equation}
	\mathcal{T}_{MD}=\|\bb{w}_{MD}^H\|^2=\frac{\bb{b}^H\bb{C}^{-2}\bb{b}}{(\bb{b}^H\bb{C}^{-1}\bb{b})^2}.
	\label{eq:101g}
\end{equation}
It should be noted that the sensitivity measure is the reciprocal of the array white-noise gain (WNG). This is another important array performance measure; it represents the ratio between the signal-to-noise ratio (SNR) at the array input to the SNR at the beamformer output, assuming spatially-white noise at the input \cite{VanTrees2002}.

Based on the expression in \eqref{eq:101g}, the robustness of the maximum-directivity beamformer and its relation to the proposed measure of array quality will be analyzed in section \ref{sec:sign} and further discussed in section \ref{sec:ex}.

\subsection{Direction of arrival estimation}
\label{sec:bdoa}
Another common array processing method is direction of arrival (DOA) estimation.
One of the widely studied DOA estimation approaches is based on the multiple signal classification (MUSIC) algorithm \cite{Schmidt1986}.  
This approach involves decomposition of the measurement space into signal and noise subspaces and is based on the eigendecomposition of the measurement covariance matrix. 

In order to assess the performance of a MUSIC-based DOA estimator we refer to the measurement model in \eqref{eq:101}. Recall that there are $D$ sources with directions $\{\Omega_j\}_{j=1}^D$ to be estimated from the measurements of $L>D$ microphones. 
In addition, it is assumed that $\bb{s}$ and $\bb{n}$ are zero mean, uncorrelated random vectors. 
Eigendecomposition of the measurement covariance matrix $\bb{P}=\mE{\bb{pp}^H}$, where $\mE{\cdot}$ denotes the expectation operator, can be written as:
\begin{equation}
	\bb{P}=\bb{Q\Lambda Q}^H,
	\label{eq:150}
\end{equation}
where the columns of $\bb{Q}=[\bb{q}_1\,\,\bb{q}_2\,\,\cdots\,\,\bb{q}_L]\in\mathbb{C}^{L\times L}$ are the eigenvectors of $\bb{P}$ and $\bb{\Lambda}=\mathrm{diag}(\lambda_1,\lambda_2,...,\lambda_L)$ is a diagonal matrix with $\{\lambda_i\}_{i=1}^L$ denoting the eigenvalues of $\bb{P}$. It is also assumed that $\{\lambda_i\}$ are arranged in descending order and
that the noise is spatially white, i.e. $\mE{\bb{nn}^H}=\sigma\bb{I}$, where $\sigma$ is the noise power at each microphone. It is further assumed that the source direction is described by a single parameter, e.g. elevation, while the azimuth remains fixed. 
The MUSIC estimator of $\Omega_j$ is given by $\hat{\Omega}_j=\mathrm{argmax}_{\Omega}\{P_{MU}(\Omega)\}$, where $P_{MU}$ is the MUSIC pseudospectrum (see e.g. \cite{Schmidt1986} for details).
The variance of $\hat{\Omega}_j$ can be asymptotically expressed by \cite{Stoica1989}:
\begin{equation}
	\mE{(\hat{\Omega}_j-\Omega_j)^2}=c\sigma\sum\limits_{i=1}^D{\frac{\lambda_i}{(\sigma-\lambda_i)^2}|\bb{a}^H(\Omega_j)\bb{q}	_i|^2},
	\label{eq:151}
\end{equation}
where $\bb{a}(\Omega_j)$ is the array steering vector for the source at direction $\Omega_j$ and $c$ is a constant that depends on the number of snapshots used for the estimation and on the derivative of $\bb{a}(\Omega)$ at $\Omega_j$ (see \cite{Stoica1989} for more details). It is assumed that $c$ is independent of $j$, which is approximately true in the case of closely spaced sources.

The expression in \eqref{eq:151} will be utilized in section \ref{sec:doa} in order to investigate the relation between the performance of the MUSIC DOA estimator and the measure of array quality that is proposed in section \ref{sec:meas}. 

	\section{Generalized measurement model}
\label{sec:sIII}
A generalized wide-band measurement model is presented in this section by extending the standard narrow-band model presented in section \ref{sec:measmod}. This wide-band model will then facilitate the development of a wide-band quality measure for the array.

Consider, as before, a humanoid robot head submerged in a sound field produced by $D$ spatially separated sources and an array consisting of $L$ microphones distributed on the head surface. The measurement model in \eqref{eq:101} can be extended to the wide-band case by concatenating the array measurements taken at $K$ different frequencies in a single column vector $\bb{p}\in\mathbb{C}^{LK\times 1}$, i.e.
\begin{equation}
	\bb{p}=[\bb{p}(\omega_1)^T\,\,\bb{p}(\omega_2)^T\,\,\cdots\,\,\bb{p}(\omega_K)^T]^T,
	\label{eq:701}
\end{equation}
where
\begin{equation}
	\bb{p}(\omega_k)=[p_1(\omega_k)\,\,p_2(\omega_k)\,\,\cdots\,\,p_L(\omega_k)]^T,\,\,\,k=1,2,...,K.
\label{eq:702}
\end{equation}
The source signal and the noise vectors can be extended in a similar way to obtain $\bb{s}\in\mathbb{C}^{DK\times 1}$ and $\bb{n}\in\mathbb{C}^{LK\times 1}$, respectively.
In this case, matrix $\bb{A}$ that relates the source signals and the measurements will be a block-diagonal matrix, with each block consisting of array steering vectors at a single frequency.

The current work aims to characterize the overall array performance combined across space and frequency.
For this purpose, the extension described above has a major drawback, because it treats frequencies independently by placing the steering vectors for different frequencies on separate columns of $\bb{A}$.
In addition, this extension involves a relatively large $LK\times DK$ matrix that can considerably increase the computational complexity of the method proposed here. 
Instead of using this straightforward extension, here it is proposed to use a model that describes the frequency response of the array to a single source at (i) a known frequency range and (ii) a known range of possible source directions.
This model overcomes both drawbacks of the above-mentioned straightforward wide-band extension; it allows to combine the measurements made by the different microphones and at different frequencies along the same dimension of the model transfer matrix, therefore reducing its dimensions. 
The proposed measurement model is given by:
\begin{equation}
		\bb{p}=\bb{Hs}+\bb{n},
	\label{eq:703}
\end{equation}
where $\bb{p}\in\mathbb{C}^{LK\times 1}$ holds the complex pressure amplitudes measured by $L$ microphones at $K$ different frequencies, as defined in \eqref{eq:701} and \eqref{eq:702} and 
vector $\bb{s}\in\mathbb{C}^{D\times 1}$ holds the amplitude of a source located in $D$ possible source directions, i.e.
\begin{equation}
	\bb{s}=[s(\Omega_1)\,\,s(\Omega_2)\,\,\cdots\,\,s(\Omega_D)]^T\in\mathbb{C}^{D\times 1}.
	\label{eq:704}
\end{equation}
Here, the model describes a frequency response and, in contrast to \eqref{eq:101}, the entries of $\bb{s}$ are not a function of frequency. The model can still be made more flexible by extending $\bb{H}$ to $\bb{W}_1\bb{HW}_2$, where $\bb{W}_1$ and $\bb{W}_2$ are diagonal weighting matrices that can provide larger weights to preferred frequencies and preferred directions, respectively. 
An additional difference between the definitions of $\bb{s}$ in equations \eqref{eq:101} and \eqref{eq:703} is that here the HRTF notation is adopted. This means that the amplitudes are not measured at the sources, but at the center of the robot head when it is removed \cite{Shaw1974}, and that the steering matrix $\bb{A}$ is replaced by the transfer function matrix $\bb{H}$, as explained shortly. The two notations are essentially the same up to a normalization constant that does not depend on the source direction and, therefore, has no effect on the methods described here. This constant will be omitted in the subsequent discussion. The HRTF notation is adopted here because it is more common in the context of human spatial hearing and, therefore, more convenient in the context of microphone arrays for humanoid robot audition.

Matrix $\bb{H}\in\mathbb{C}^{LK\times D}$ consists of the generalized head related transfer functions (GHRTFs) that describe the response at any given position on the head surface. The word \emph{generalized} is used here to emphasize the fact that the HRTFs  are not limited to the left or the right ears, as is common in the human spatial hearing literature. In particular, the  elements $h_{ij}$ of the GHRTF matrix $\bb{H}$ describe the transfer function between the source indexed by $j$ and the microphone and frequency indexed by $i$, i.e.
\begin{equation}
	p_l(\omega_k)=h_{ij}\cdot s(\Omega_j),
	\label{eq:1014}
\end{equation}
while index $i$ is a function of both $l$ and $k$ and is given by 
\begin{equation}
	i=L\cdot (k-1)+l, 
	\label{eq:1015}
\end{equation}
with $l=1,2,...,L$ and $k=1,2,...,K$. 
The operating frequency range of interest and the expected range of source directions can be incorporated by an appropriate selection of the ranges of indices $i$ and $j$. 
Finally, $\bb{n}\in\mathbb{C}^{KL\times 1}$ represents noise, either present in the sound field or added by the instrumentation, at all $K$ frequencies of interest.
It is emphasized that in the narrow-band case, i.e. for $K=1$, we obtain $i=1,2,...,L$. Thus, $i$ simply represents the microphone index and the GHRTF matrix $\bb{H}$ reduces to the array steering matrix $\bb{A}$. Hence, in the narrow-band case the generalized model described here reduces to the model presented in \eqref{eq:101}.
	\section{Effective Rank - A Measure of Array Quality}
\label{sec:meas}
The goal of the work presented here is to develop a measure of array quality as a function of microphone positioning that will enable optimization of microphone placement.
In this section, the noise vector $\bb{n}$ is ignored, i.e. it is assumed that $\bb{n}=\bb{0}$. This is in order to concentrate on the development of a measure that is dependent on the GHRTF matrix $\bb{H}$ and not on the SNR or other mutual signal and noise properties.

The measure of array quality is developed by quantifying the information in the measurement $\bb{p}$ related to the arriving signal amplitudes $\bb{s}$ as a function of the GHRTF matrix $\bb{H}$.
For this purpose, consider the singular value decomposition (SVD)\cite{Golub1996} of the GHRTF matrix $\bb{H}$:
\begin{equation}
	\bb{H}=\bb{U\Sigma V}^H,
\label{eq:102}
\end{equation}
where $\bb{\Sigma}=\mathrm{diag}\{\sigma_1,\sigma_2,...,\sigma_{\min(KL,D)}\}\in\mathbb{C}^{KL\times D}$ is a diagonal matrix consisting of the singular values of $\bb{H}$, which are positive real scalars arranged in descending order:
$\sigma_1\geq\sigma_2\geq\sigma_q\geq\sigma_{q+1}=\cdots=\sigma_{\min(KL,D)}=0$, with $q=\mathrm{rank}\{\bb{H}\}$.
Note that in most practical cases $q$ will be equal to $\mathrm{min}(KL,D)$, because, in practice, it is not likely to obtain singular values that are exactly zeros. Matrices $\bb{U}=[\bb{u}_1\,\,\bb{u}_2\,\,\cdots\,\,\bb{u}_{KL}]\in\mathbb{C}^{KL\times KL}$ and $\bb{V}=[\bb{v}_1,\,\,\bb{v}_2\,\,\cdots\,\,\bb{v}_D]\in\mathbb{C}^{D\times D}$ consist of orthogonal columns, which are the left and the right singular vectors of $\bb{H}$, respectively. 
The vectors $\{\bb{u}_1,\bb{u}_2,...,\bb{u}_q\}$ span the range space of $\bb{H}$, such that the measured pressure $\bb{p}$ is given by:
\begin{align}
	\nonumber\bb{p}&=\bb{U\Sigma V}^H\bb{s}\\
	&=\sum\limits_{i=1}^{q}{c_i\cdot \bb{u}_i},
	\label{eq:103}
\end{align}
where 
\begin{equation}
	c_i=\sigma_i\cdot\bb{v}_i^H\bb{s},\,\,\,i=1,2,...,q. 
	\label{eq:104}
\end{equation}
Equations \eqref{eq:103} and \eqref{eq:104} imply that regardless of the number of frequencies $K$ and the number of microphones $L$ at which the pressure is measured, the number of the elements of $\bb{s}$ that can be reconstructed from $\bb{p}$ given $\bb{H}$ is at most equal to the rank of $\bb{H}$.  

Inspired by the above observation, it seems natural to define the amount of information for the reconstruction of $\bb{s}$ from $\bb{p}$ as the rank of the GHRTF matrix $\bb{H}$. However, the strict definition of matrix rank may not provide an adequate measure. 
To illustrate this, consider a matrix $\bb{H}$ with $q$ non-zero singular values, but with the last $q-r$ singular values having negligibly small values compared to the first $r$ singular values, where $1\leq r< q$. In this case, the range space of $\bb{H}$ is effectively spanned by $\{\bb{u}_1,\bb{u}_2,...,\bb{u}_r\}$, although $\mathrm{rank}\{\bb{H}\}=q$. 
The effective dimension $r$ of the range space of $\bb{H}$ can be estimated using the techniques for the estimation of the dimension of the signal subspace reported in the literature \cite{Wax1985,Fishler1999}. However, the solutions presented there involve assumptions on the statistics of the measurements and yield only an integer estimate. This can be limiting in practice for several reasons, including: (i) two different GHRTF matrices having the same estimated integer effective range space dimension may still result in considerable difference in array performance, (ii) array optimization algorithms may result in a local minimum without being able to proceed, when based on an integer array quality measure.    

In order to overcome these problems we propose to base the measure of array quality on the \emph{effective rank} of the GHRTF matrix \cite{Roy2007}. In particular, given $\bb{H}$ with $q$ non-zero singular values, the effective rank of $\bb{H}$ is defined as follows:
\begin{equation}
	\mc{R}(\bb{H})=\exp\left(-\sum\limits_{i=1}^{q}{\bar{\sigma}_i\cdot\log{\bar{\sigma}_i}}\right),
	\label{eq:105}
\end{equation}
where
\begin{equation}
	\bar{\sigma}_i=\frac{\sigma_i}{\sum\limits_{j=1}^{q}{\sigma_j}},
\label{eq:105a}
\end{equation}
with $\{\sigma_j\}$ and $q$ denoting the singular values and the rank of $\bb{H}$, respectively, as above (see \eqref{eq:102}), and $\log(\cdot)$ is the natural logarithm. Note that the argument of the exponential in \eqref{eq:105} is the Shannon's entropy of a discrete random variable with a probability mass function described by $\{\bar{\sigma}_i\}$ \cite{Cover2006}. The entropy is bounded by:
\begin{equation}
	0\leq -\sum\limits_{i=1}^{q}{\bar{\sigma}_i\cdot\log{\bar{\sigma}_i}} \leq \log q,
	\label{eq:106}
\end{equation}
implying that the effective rank $\mathcal{R}(\bb{H})$ is a real positive number limited by:
\begin{equation}
	1 \leq\mathcal{R}(\bb{H})\leq q.
	\label{eq:107}
\end{equation}
Effective rank measures the uniformity of the singular values of $\bb{H}$. It is maximized when all singular values are equal, i.e. $\sigma_j=\sigma_1,\,\,j=1,2,...,q$. In this case the effective rank and the rank of $\bb{H}$ are equal. Effective rank is minimized when the first singular value is much larger than the other singular values, i.e. $\sigma_1>>\sigma_j,\,\,j=2,3,...,q$. In this case the effective rank of the matrix is near unity.

	\section{Optimal Microphone Positioning and Array Performance}
\label{sec:sign}
In this section the relations between the effective rank and array performance are analyzed. This is done by first defining the optimal microphone positioning problem using the generalized model presented above and the proposed array quality measure. Then, the effective rank of the array is related to performance with beamforming and DOA estimation algorithms.

\subsection{Optimal microphone positioning} 
\label{sec:optarr}
The optimal microphone positioning problem can be formulated as a selection problem, where it is required to choose $L$ positions out of a discrete set of a total of $M$ possible positions on the head surface. Consider the set of all possible microphone positions 
$\mc{M}$ on a humanoid robot head surface. Denote the GHRTF matrix at position $m\in\mc{M}$ by $\bb{H}_m$. The optimization problem can be stated as follows:
\begin{equation}
	\mc{L}^{\star}=\mathop{\mathrm{argmax}}\limits_{\mc{L}}\mc{R}(\bb{H}_{\mc{L}}),
	\label{eq:51}
\end{equation}
where $\mc{L}=\{l_1,l_2,...,l_L\}\subset\mc{M}$ is the subset to be chosen, and $\bb{H}_{\mc{L}}=\left[\bb{H}_{l_1}^T\,
\bb{H}_{l_2}^T\,\cdots\,\bb{H}_{l_L}^T\right]^T$ is the column concatenation of the GHRTF matrices of all positions in the subset.

Direct solution to the problem in \eqref{eq:51} requires the evaluation of each of $\frac{M!}{L!(M-L)!}$ possible combinations, which is not realistic using current machine computation abilities for practical $M$ and $L$. However, the solution can be obtained using the Genetic Algorithms (GA) optimization approach (for a review see e.g. \cite{Weile1997}), as demonstrated in section \ref{sec:ex}. While this approach is based on a heuristic and optimality of its results is not guaranteed, it has been successfully applied in the past decades to a wide range of optimization problems related to antenna \cite{Weile1997} and microphone \cite{Pessentheiner2012} array design. 

In the following two subsections the advantage of increasing the effective rank of the array when used with beamforming and DOA estimation algorithms is analyzed. It is assumed that increasing the effective rank of $\bb{H}$ in the wide-band case will increase the effective rank of the GHRTF matrix at each frequency separately. Based on this assumption, the analysis here is performed in the narrow-band case, i.e. for $K=1$ and $\bb{A}=\bb{H}$.

\subsection{Relation to Beamformer Robustness} 
\label{sec:beamf}
The complex pressure amplitudes $\bb{p}$ measured on the humanoid robot head surface can be processed to perform various tasks, including spatial filtering. As mentioned above, the robustness of a beamfomer can be assessed through the beamformer sensitivity measure\cite{VanTrees2002}. 
An expression for the sensitivity of the maximum-directivity beamformer is given in \eqref{eq:101g}. By substituting into \eqref{eq:101g} the approximation of $\bb{C}$ given in \eqref{eq:101e} with the notation of the generalized model, i.e. $\bb{C}\approx\frac{1}{D}\bb{HH}^H$, we obtain:
\begin{equation}
	\mc{T}_{MD}=\frac{\bb{b}^H\left(\bb{HH}^H\right)^{-2}\bb{b}}{\left(\bb{b}^H(\bb{HH}^H)^{-1}\bb{b}\right)^2}.
	\label{eq:203}
\end{equation}
Thus, the sensitivity of the maximum-directivity beamformer depends on the GHRTF matrix $\bb{H}$, which, in turn, depends on microphone positioning.
By using the SVD of $\bb{H}$, i.e. $\bb{H}=\bb{U\Sigma V}^H$, we obtain:
\begin{equation}
	\mathcal{T}_{MD}=\frac{\bb{b}^H\bb{U\Sigma}^{-4}\bb{U}^H\bb{b}}{(\bb{b}^H\bb{U\Sigma}^{-2}\bb{U}^H\bb{b})^2}
	=\frac{\sum\limits_{i=1}^{q}{\frac{1}{\sigma_i^4}|\bb{u}_i^H\bb{b}|^2}}
	{\left(\sum\limits_{i=1}^{q}{\frac{1}{\sigma_i^2}|\bb{u}_i^H\bb{b}|^2}\right)^2},
	\label{eq:204}
\end{equation}
where $\bb{\Sigma}^{-2}=\left(\bb{\Sigma\Sigma}^H\right)^{-1}$ and it is assumed that $\bb{H}$ has full rank, i.e. $q=L$.
Now, recall that maximizing the effective rank of $\bb{H}$ will tend to produce a more uniform distribution of its singular values. In the limiting case, where all singular values are equal, i.e. $\sigma_i=\sigma_1,\,\,i=2,...,q$, the expression in \eqref{eq:204} reduces to 
\begin{equation}
	\mathcal{T}_{MD}=\frac{1}{\sum\limits_{i=1}^{q}{|\bb{u}_i^H\bb{b}|^2}}
	=\frac{1}{\bb{b}^H\bb{UU}^H\bb{b}}
	=\frac{1}{\bb{b}^H\bb{b}}.
	\label{eq:205}
\end{equation}
The resulting expression in \eqref{eq:205} is familiar in the array processing literature\cite{VanTrees2002}; it represents the lower bound on sensitivity for the maximum-directivity beamformer when constrained to a distortionless response at a given look direction characterized by $\bb{b}$. Thus, an array with maximum effective rank attains minimum achievable sensitivity.
This result implies that maximizing the effective rank of the GHRTF matrix will generally tend to reduce the sensitivity and therefore improve the robustness of the maximum-directivity beamformer.

\subsection{Relation to DOA estimation accuracy} 
\label{sec:doa}
The relation between the proposed measure and the performance of the MUSIC DOA estimator is analyzed in this subsection. Recall that the analysis presented in this section is based on the notation of the generalized measurement model \eqref{eq:703} in the narrow-band case i.e. for $K=1$ and $\bb{H}=\bb{A}$.

Here, in addition to the assumptions made in section \ref{sec:bdoa}, it is assumed that the sources are uncorrelated and have the same power $\alpha$, i.e. $\mathrm{E}[\bb{ss}^H]=\alpha \bb{I}$. In this case, the covariance matrix of the measurements $\bb{p}$ is given by:
\begin{equation}
	\mathrm{E}[\bb{pp}^H]=\alpha\bb{HH}^H+\sigma\bb{I}=\alpha\bb{U\Sigma}^2\bb{U}^H+\sigma\bb{I}
	=\bb{U\Lambda U}^H,
	\label{eq:210}
\end{equation}
where $\bb{U}$ and $\bb{\Sigma}$ contain the left singular vectors and the singular values of $\bb{H}$, as above. 
Matrix $\bb{\Lambda}=\mathrm{diag}\{\lambda_1,\lambda_2,...,\lambda_L\}$ holds the eigenvalues of $\mE{\bb{pp}^H}$, as in \eqref{eq:150}. From \eqref{eq:210}, the eigenvalues $\{\lambda_i\}$ are given by:
\begin{equation}
	\lambda_i=\left\{\begin{tabular}{ll}
							$\alpha\sigma_i^2+\sigma$, & $i=1,2,...,D$ \\
							$\sigma$, & $i=D+1,...,L$ 
						\end{tabular}\right.,
	\label{eq:211}
\end{equation}
where $\{\sigma_i\}$ denote the singular values of $\bb{H}$, as above. 
By substituting the results in \eqref{eq:210} and \eqref{eq:211} into the expression for the variance of the MUSIC DOA estimator of the arrival direction $\Omega_j$ given in \eqref{eq:151}, we obtain: 
\begin{equation}
	\mathrm{E}\left[(\hat{\Omega}_j-\Omega_j)^2\right]
	=\frac{c\sigma}{\alpha}{\sum\limits_{i=1}^{D}{\frac{\sigma_i^2+\frac{\sigma}{\alpha}}
	{\sigma_i^4}|\bb{h}_j^H\bb{u}_i|^2}},
	\label{eq:212}
\end{equation}
where $\hat{\Omega}_j$ is the MUSIC estimator of $\Omega_j$, $\bb{h}_j$ is the $j^{th}$ column of $\bb{H}$, being the steering vector of the array for the direction $\Omega_j$, and $\{\bb{u}_i\}_{i=1}^{D}$ are the first $D$ left singular vectors of $\bb{H}$, which, from \eqref{eq:210}, are also the eigenvectors of $\mE{\bb{pp}^H}$. By summing estimation error variances over all $D$ sources, we obtain:
\begin{align}
	\nonumber\sum\limits_{j=1}^{D}{\mathrm{E}\left[(\hat{\Omega}_j-\Omega_j)^2\right]}
	&=\frac{c\sigma}{\alpha}\sum\limits_{i=1}^{D}{\frac{\sigma_i^2+\frac{\sigma}{\alpha}}{\sigma_i^4}
	\sum\limits_{j=1}^{D}{|\bb{h}_j^H\bb{u}_i|^2}}\\
  \nonumber&=\frac{c\sigma}{\alpha}\sum\limits_{i=1}^{D}{\frac{\sigma_i^2+\frac{\sigma}
  {\alpha}}{\sigma_i^4}\|\bb{H}^H\bb{u}_i\|^2}\\
  &=\frac{cD}{\mathrm{SNR}}+\frac{c}{{\mathrm{SNR}}^2}\sum\limits_{i=1}^{D}{\frac{1}{\sigma_i^2}},
	\label{eq:213}
\end{align}
where $\mathrm{SNR}=\alpha/\sigma$. Only the second term on the right-hand side of \eqref{eq:213} depends on the singular values of $\bb{H}$. Thus, for high $\mathrm{{SNR}}$ the singular values have no effect on the total estimation error. However, for low $\mathrm{{SNR}}$ values the second term in \eqref{eq:213} is dominant. 
By considering the constraint of constant matrix gain, i.e. $\sum_{i=1}^D{\sigma_i^2}=const$, it can be shown using the Lagrange multipliers method that the second factor, i.e. $\sum_{i=1}^{D}{1/\sigma_i^2}$ is minimized when $\sigma_i=\sqrt{const/D},\,i=1,2,...,D$. This implies that the estimation error is minimized for uniform singular values.
Thus, increasing the effective rank of $\bb{H}$ will generally tend to decrease the MUSIC DOA estimation variance under the above assumptions, especially for low $\mathrm{{SNR}}$ values. 

Note that \eqref{eq:213} implies that there is an improvement in the performance at each frequency independently. There are various possible extensions of the MUSIC algorithm to the wide-band case. One possible approach is to combine the narrow-band estimates to produce a single wide-band result. Another approach is related to beamspace based techniques \cite{Ward2004,Khaykin2009}. These techniques are based on a field representation that allows the separation between frequency and space dependent parts. This field representation is first estimated at each frequency independently. Then, a final estimate is obtained by averaging the frequency dependent part of the covariance matrix and performing a narrow-band estimation. Either way, the above-mentioned techniques for the wide-band extension of MUSIC rely to some extent on the narrow-band estimates and, therefore, are expected to improve when the narrow-band performance is improved.

	\section{Simulation Study} 
\label{sec:ex}
This section provides an example of array design using the framework derived in the previous sections. First, a method for obtaining the GHRTF database is described. Then, optimal microphone positioning and its advantages in microphone array processing are demonstrated.

\subsection{Computation of GHRTFs from head geometry} 
\label{sec:hrtfdb}
In order to perform a numerical study using the proposed theoretical framework, a GHRTF database is required. One such database is available from \cite{Maazaoui2012}, but it contains the GHRTFs for only $16$ different positions over the head and for a limited range of arrival directions. In this section, a method for generating a more comprehensive GHRTF database is described.

The database was obtained by means of a numerical simulation using the boundary element method (BEM). The algorithm was implemented in MATLAB\circledR~based on the method described in \cite{Chandler2007}. In this implementation the entire head surface is divided into small sub-surfaces \-- the elements. Then, the sound pressure on the surface of the head is approximated by assuming a constant value at each element. Finally, the pressure at each element is found by solving the Helmholtz equation at each frequency of interest. The well known problem of non-unique solutions at irregular frequencies (see, for example \cite{Wrobel2002}, chap. 5) was resolved using the CHIEF method \cite{Schenck1968}.

The algorithm was validated by calculating the HRTFs for the KU-$100$ dummy head surface. The geometry of the surface was kindly provided by Brian F. G. Katz, who used this model for the study reported in \cite{Katz2007}. For validation purposes a relatively detailed mesh was used, consisting of about $37000$ elements. The calculation of HRTFs was limited to $9$ kHz due to the required computational effort. The results obtained using the current algorithm were, on average, within $1$ dB from the HRTFs available in the literature for the same geometry model \cite{Katz2007}.

\begin{figure} 
	\centering
	\includegraphics[width=0.4\columnwidth]{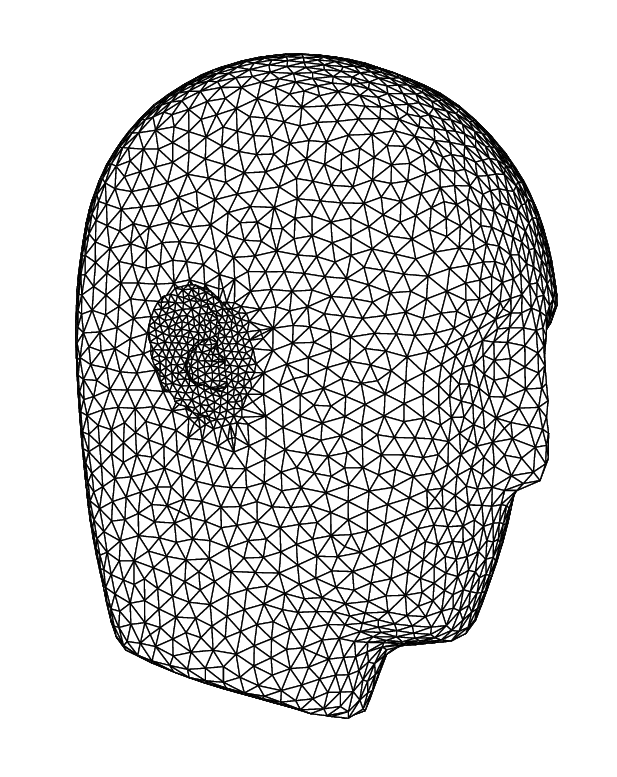}
	\caption{The dummy head geometry model that was used for the construction of the GHRTF database.}
	\label{fig:1}
\end{figure}

Construction of the GHRTFs database involves calculation of the GHRTFs at a relatively large number of positions on the head and as a response to a large number of different source positions. To make these calculations computationally feasible, the KU-$100$ dummy head geometry was simplified by removing the fine details of the neck joint and smoothing sharp edges. The resulting model is presented in Fig. \ref{fig:1}. Although simplified, it still contains several important parts of a human head geometry including ears, nose, eye sockets and chin, and can serve as a good example of a humanoid robot head geometry.  
The geometry consists of $6872$ elements, with the largest element edge length being about $11$ mm. This is expected to yield a valid pressure calculation up to $5$ kHz according to the ``$6$ elements per wavelength rule'' \cite{Katz2007}. 

It was assumed that the head is acoustically rigid, based on the results reported in \cite{Katz2000}. 
The sound pressure was calculated at each of the $6872$ geometry elements for $312$ different source directions, of which $36$ were distributed uniformly in azimuth on the horizontal plane, another $36$ were distributed uniformly in elevation on the median plane, and the last $240$ were nearly uniformly distributed on the surface of a unit sphere \cite{HardinSloane1996}. For an illustration of the various source positions see the right column of Fig. \ref{fig:2}. Three different sets of directions were considered in order to illustrate the effect of a preferred set of source directions on the array design. For example, the first set is appropriate for a scenario where all sources of interest are located at the level of the robot's head. In addition, having the third set of $240$ nearly uniformly distributed source directions, it is possible to select any desired subset to reflect a specific scenario of practical importance. The obtained database consists of $6872$ matrices with dimensions of $51\times 312$. These matrices represent the generalized HRTFs for $6872$ microphone positions on the head surface at the frequencies $\{0, 100, 200, ..., 5000\}$ Hz and for $312$ different source positions. A linear frequency scale was selected here because this work focuses on the HRTFs. These are typically represented by the discrete Fourier transforms (DFTs) of the head related impulse responses (HRIRs) and therefore use a  linear frequency scale. Nevertheless, e.g. for speech processing, it may be possible to use a logarithmic frequency scale by selecting a logarithmically spaced subset from the current GHRTF database.

\subsection{Relation between the geometry and the effective rank} 
Here, the relation between a microphone position on the head surface, its associated GHRTF and the effective rank are analyzed.
The aim of this analysis is to gain an insight into the way in which head surface geometry affects array performance through the effective rank measure.

\begin{figure}[ht] 
	\centering
	\includegraphics[width=\imgwidth\columnwidth]{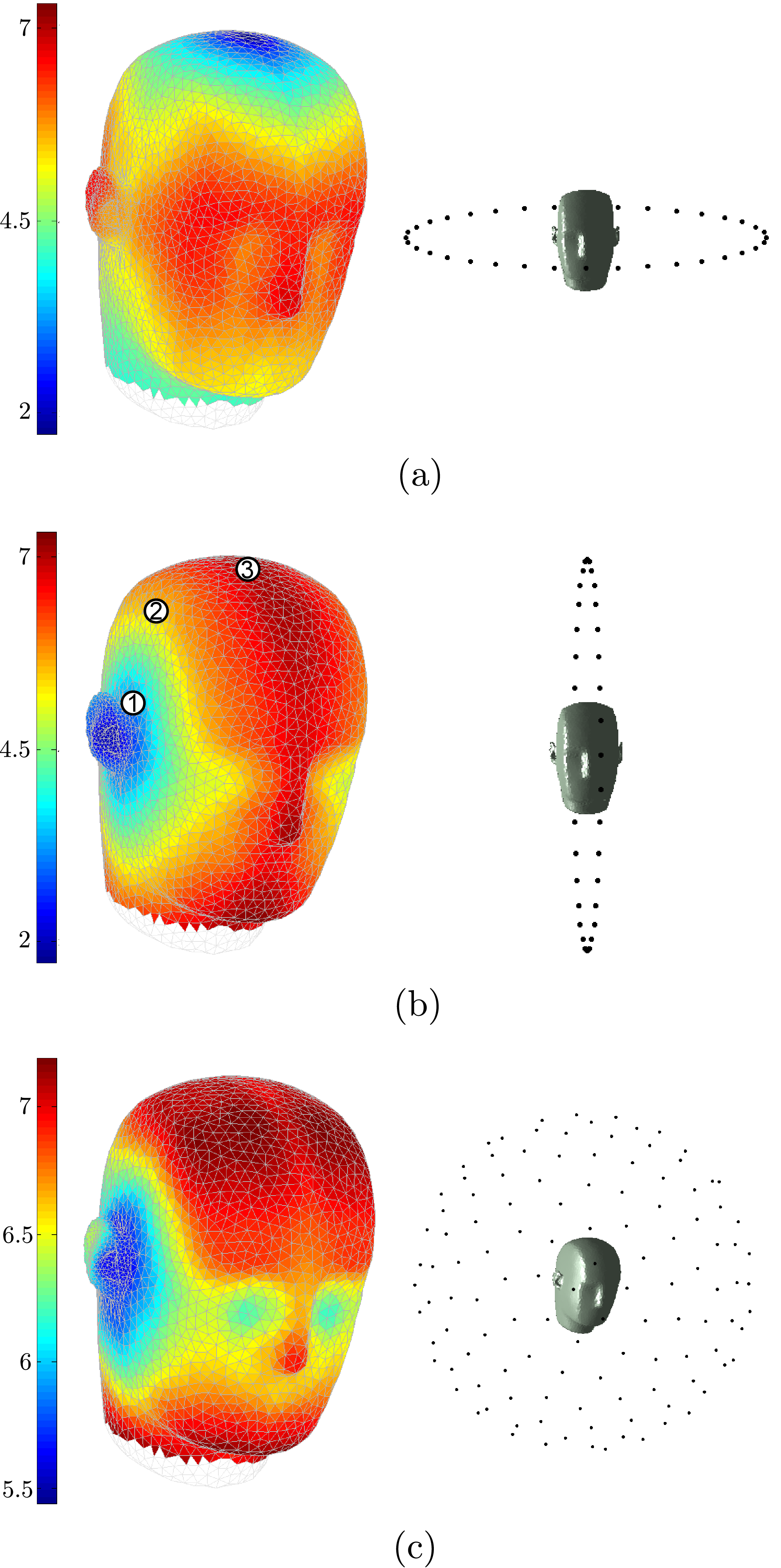}
	\caption{Effective rank calculated for three different source arrangements: (a) \-- sources on horizontal plane, (b) \-- sources on
	median plane, (c) \-- nearly uniformly distributed sources. Left column represents the effective rank, right column illustrates
	the	source distribution.}
	\label{fig:2}
\end{figure}

First, different positions on the head surface are rated for their appropriateness for microphone positioning. This is done by calculating the effective rank of an array consisting of a single microphone, which is located at one given position at a time.
Recall that each matrix in the GHRTF database represents $\bb{H}$ (see the model in \eqref{eq:703}) for a given microphone position that describes the response at $51$ different frequencies to $312$ different source directions. The effective rank at each position was calculated in three different configurations that differed from one another in the source directions that were taken into account: (i) \-- $36$ directions uniformly distributed on the horizontal plane, (ii) \-- $36$ directions uniformly distributed on the median plane, (iii) \-- $240$ nearly uniformly \cite{HardinSloane1996} distributed directions. The results are plotted in Fig. \ref{fig:2}.
It can be seen that for vertically and horizontally distributed sources the effective rank is higher on vertical and horizontal strips, respectively. In the third configuration, i.e. $240$ nearly uniform source directions, the effective rank is higher on the upper and lower parts of the head surface. Note that the range of effective ranks in the third configuration is relatively small, as compared to the first two configurations. This could imply that when a uniform source distribution is considered the relative importance of different microphone positions is somehow decreased.

\begin{figure}[ht] 
	\centering
	\includegraphics[width=\imgwidth\columnwidth]{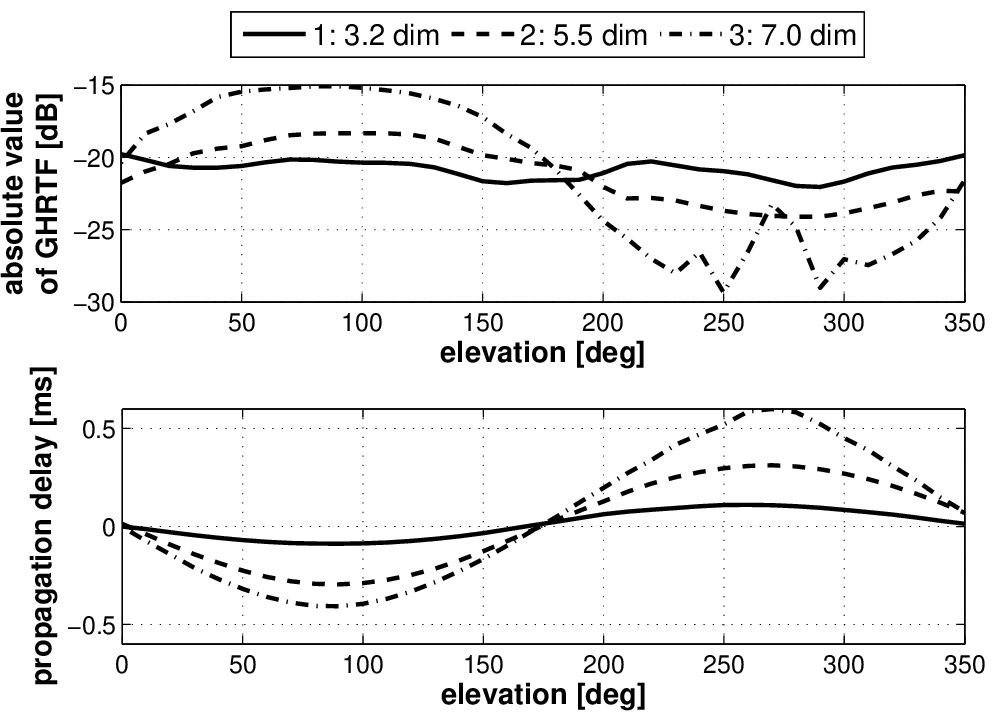}
	\caption{Plot of the GHRTFs at the frequency of $3$ kHz for the points $1,2$ and $3$ (see Fig. \ref{fig:2}b). The points, along
	with the associated effective ranks (measured in dimentions - [dim]), are indicated in the legend. For convenience, the phase
	component is plotted as the propagation delay relative to the propagation from zero elevation.}
	\label{fig:3}
\end{figure}

The aim of the next simulation is to gain insight into the mechanism behind these results. Consider the three positions indicated in Fig. \ref{fig:2}b. The GHRTFs of these points at a frequency of $3$ kHz are plotted in Fig. \ref{fig:3} as a function of elevation; the associated effective ranks are also indicated.
Note that following from point $1$ through point $2$ to point $3$ increases the effective rank. This is consistent with the increasing dependence of the GHRTFs on the direction of arrival when going from point $1$ to point $3$. This effect is noticeable for both the gain and the phase, while the gain dependence increases due to the ``shadow effect'' \cite{Blauert1997} and the phase dependence increases due to the increasing distance from the interaural axis. In addition, recall that the transfer functions at the entrance to the ear canal depend on the source elevation at higher frequencies (above $4$ kHz \cite{Blauert1997}). Hence, the effective rank of the area around the ear canal in \ref{fig:2}b, where the sources are distributed in elevation, is expected to be higher. We believe that the low effective rank in this area is due to the limited frequency range, $0-5$ kHz, that was used in this study.

To conclude, the quality of different positions on the head surface depends strongly on the source positions of interest. The quality of different positions in the frequency range considered here ($0-5$ kHz) is mostly determined by the sphere-like shape of the head which induce the propagation time differences and the ``shadow effect''.

\subsection{Sensor positioning example} 
\begin{figure}[ht] 
	\centering
	\includegraphics[width=\imgwidth\columnwidth]{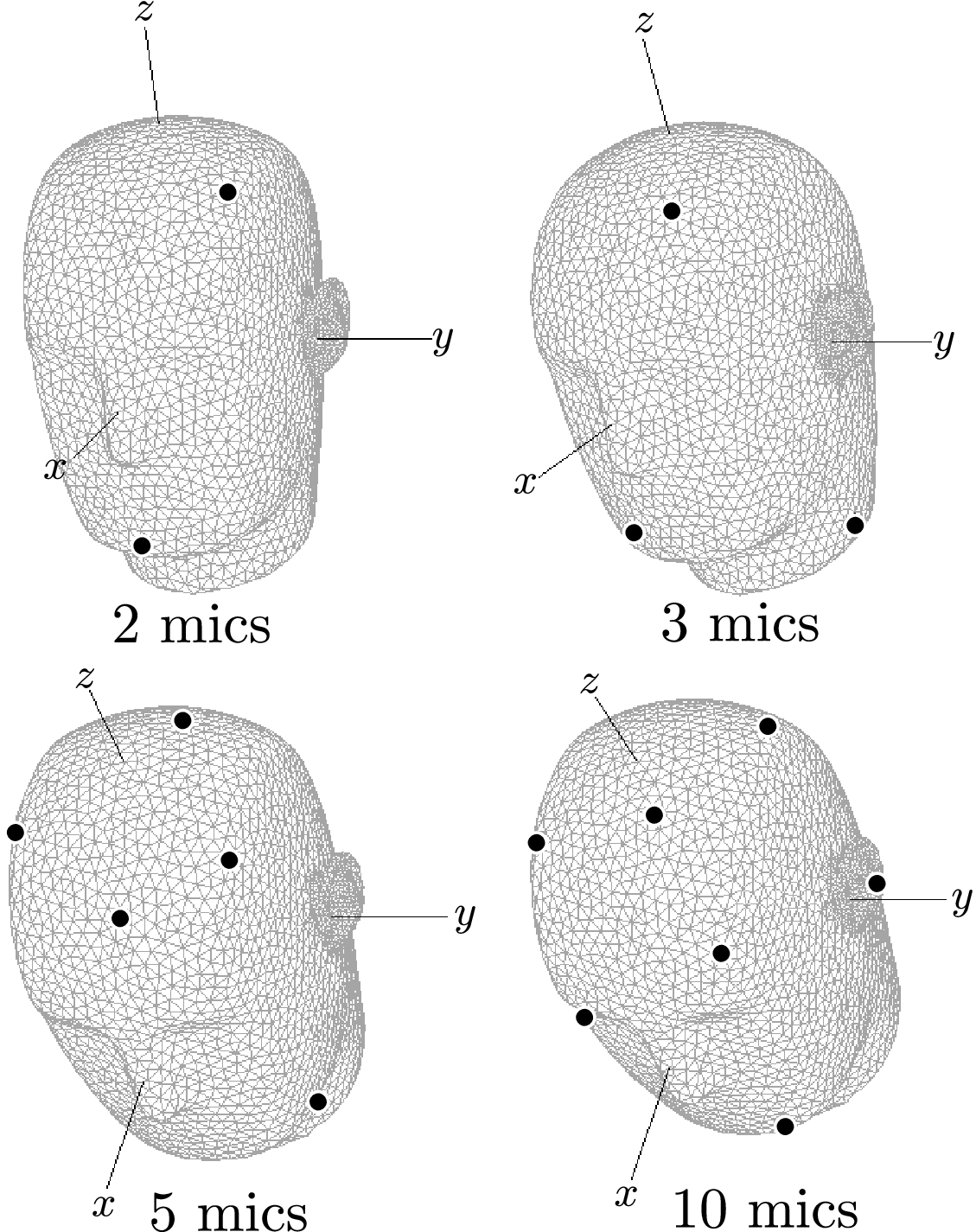}
	\caption{Representative examples of optimal positioning for $2,3,5$ and $10$ microphone arrays, obtained using the GA approach.
	Microphone positions are indicated by the black circles.}
	\label{fig:4}
\end{figure}

In this section, the problem of optimal sensor positioning defined in section \ref{sec:optarr} is solved for the KU-$100$ dummy head surface using the GHRTF database described in section \ref{sec:hrtfdb}. The solution is obtained by means of the GA based approach using the MATLAB\circledR~2012a built-in function \texttt{ga()} with the default parameters. This algorithm uses a population size of $20$ and terminates when the average change in the fitness value over $50$ subsequent generations is less than $10^{-6}$. The fitness function was $-\mc{R}(\bb{H})$, because the algorithm is designed to seek a minimum.
The optimal positioning problem was solved for the general configuration of $240$ nearly uniformly distributed sources.
Representative examples of optimal positioning for $2,3,5$ and $10$ microphones in an array are presented in Fig. \ref{fig:4}. 
Due to the stochastic nature of the GA approach, solutions were computed $100$ times for each array size. 
From the examples for $2$ and $3$ microphones it can be seen that, in general, the microphones are positioned well apart. In addition, it seems that symmetric positioning is avoided as compared, for example, to the positioning of the ears. This is probably due to the fact that symmetric positioning will increase dependence between the acquired signals and, therefore, reduce the information acquired by the sensors. As was mentioned in the introduction, the problem of optimal positioning of two microphones was solved in \cite{Skaf2011} and, in contrast to the current findings, a symmetric solution was obtained. This is because in \cite{Skaf2011} a symmetric positioning on both sides of the head was assumed a priori and the problem was actually reduced to the positioning of a single sensor. From the examples for a larger number of microphones ($5$ and $10$), it can be seen that regions with higher effective rank, i.e. the upper and lower parts of the head (see Fig. \ref{fig:2}c), are occupied more densely with sensors, while the sensors are still kept apart. For example, for the $5$-microphone array there is no microphone positioned on the sides of the head, which is a region with relatively low effective rank. However, for the $10$-microphone array, there is a microphone on the edge of the ear; this position has only an average effective rank, but allows to keep this sensor well apart from the others.  

In general, it can be concluded that the optimal positioning is achieved by a trade-off between the effective rank of individual sensor positions and their spatial separation.

\subsection{Effective rank and array robustness} 
In this section the relation between the array effective rank and its sensitivity are studied numerically. The sensitivity $\mc{T}_{maxDI}^j$ of a maximum directivity array that is constrained to a distortionless response in the look direction represented by the $j^{th}$ column of the GHRTF matrix $\bb{h}_j$ is given by Eq. \eqref{eq:203} with $\bb{b}=\bb{h}_j$. The average array sensitivity can be calculated by averaging the sensitivity over $240$ uniformly distributed look directions, which are available in the current GHRTF database:
\begin{equation}
	\bar{\mc{T}}=\frac{1}{240}\sum\limits_{j=1}^{240}{\mc{T}_{maxDI}^j}.
	\label{eq:52}
\end{equation}
In order to evaluate the effect of the effective rank of the array GHRTF matrix on the array sensitivity, the average sensitivity was calculated for four different types of array: (i) maximum effective rank (MER) array obtained in the previous section, (ii) array having an effective rank of $1$ dimension less than the MER array, (iii) array having effective rank of $5$ dimensions less than MER array, (iv) random microphone positions chosen uniformly from the available set. The last configuration represents an array design obtained without any specific engineering constraint. Note that the arrays of type (ii) and (iii) should not be designed in practice, but they will serve here in order to analyze the effect of decreased effective rank on the array performance. Solutions for the arrays of type (ii) and (iii) were obtained using the GA optimization solver. For each array type the sensitivity was  obtained as a function of frequency and of the number of microphones averaged over $100$ different array realizations. 

\begin{figure}[ht] 
	\centering
	\begin{tabular}{c}
		\includegraphics[width=\imgwidth\columnwidth]{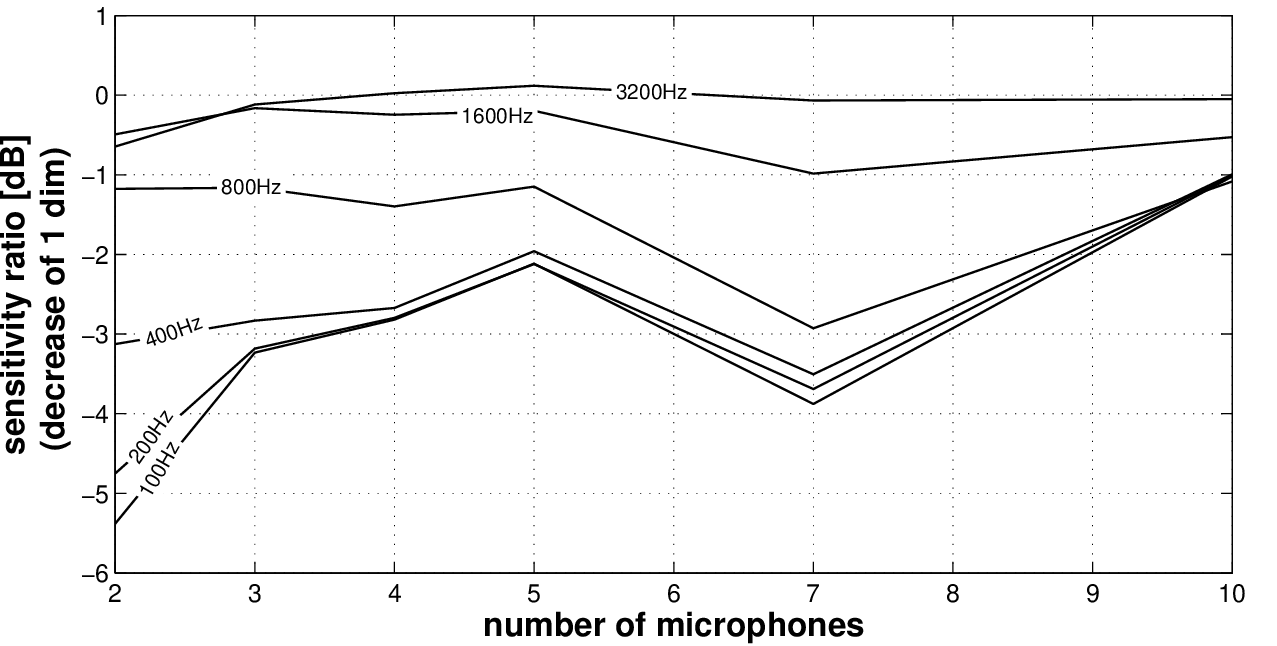}\\
		(a)\\
		\includegraphics[width=\imgwidth\columnwidth]{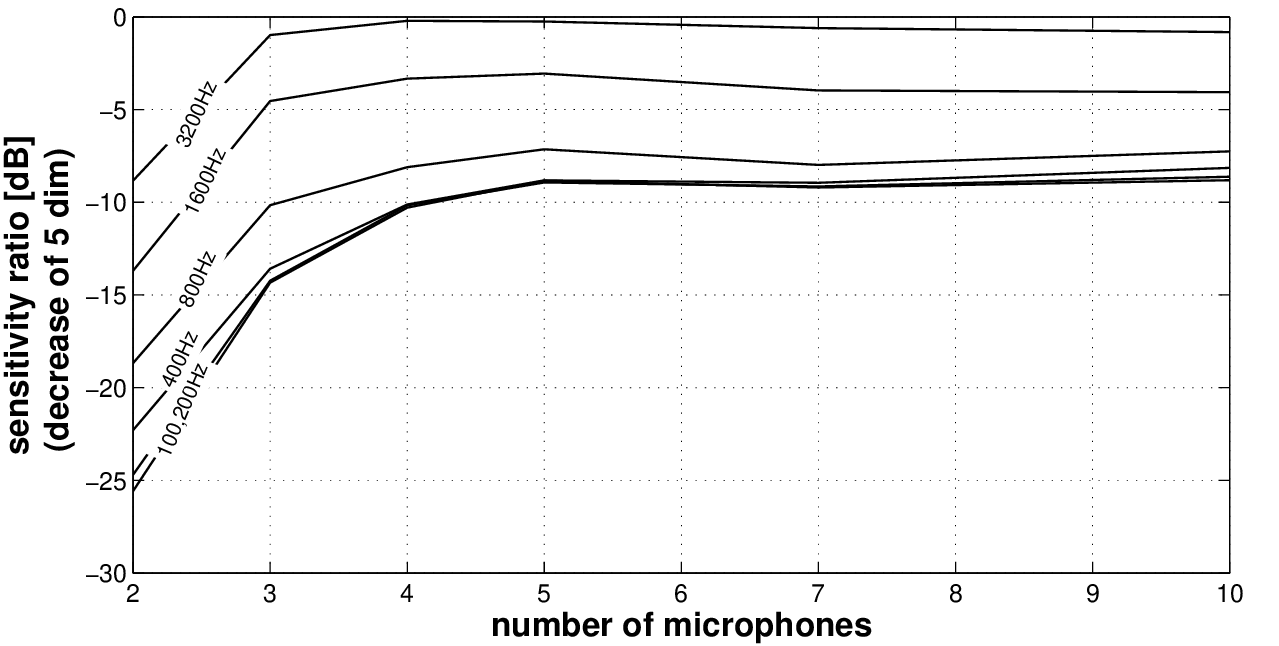}\\
		(b)
	\end{tabular}
	\caption{The ratio between the sensitivity of the MER array and the sensitivity of an
	array designed to have an effective rank that is smaller than the maximum effective rank by: (a) 1 dimension, (b) 5 dimensions.}
	\label{fig:5}
\end{figure}

The ratios of the sensitivity of the MER array to that of the arrays with decreased effective rank (types (ii) and (iii)) are shown in Fig. \ref{fig:5}.
It can be seen that, in general, increasing the effective rank of an array improves its robustness, regardless of frequency and array size. The improvement is particularly significant for smaller arrays and at lower frequencies, reaching $5$ dB and $25$ dB when the effective rank is improved by $1$ and $5$ dimensions, respectively.

\begin{figure}[ht] 
	\centering
	\begin{tabular}{c}
		\includegraphics[width=\imgwidth\columnwidth]{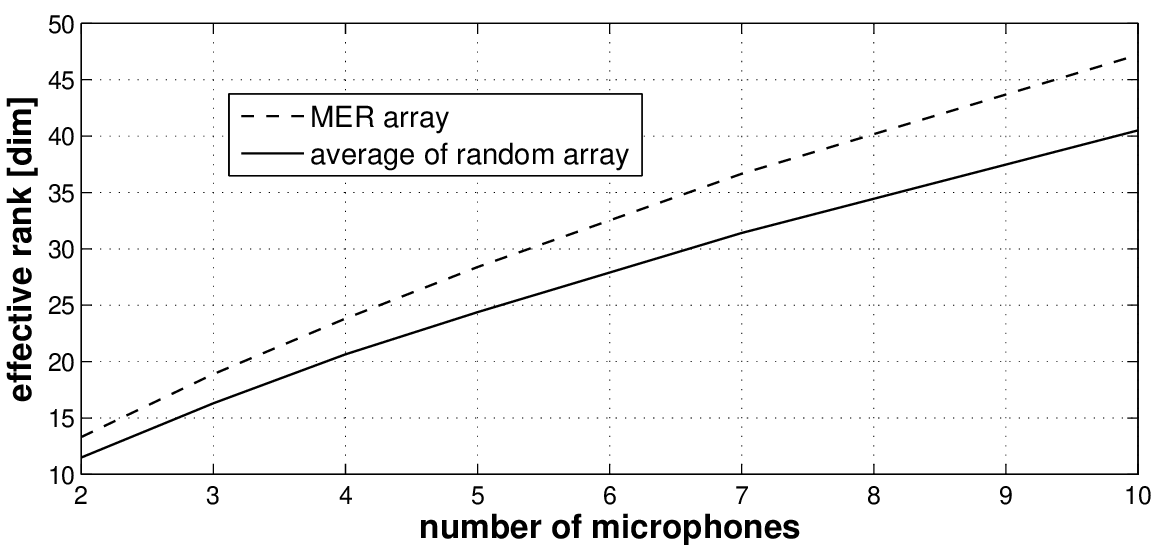}\\
		(a)\\
		\includegraphics[width=\imgwidth\columnwidth]{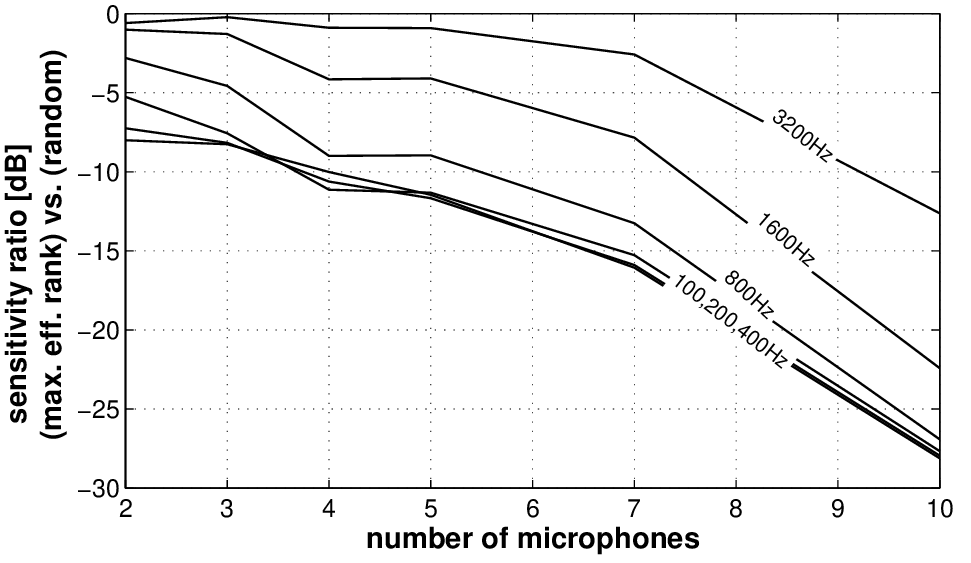}\\
		(b)
	\end{tabular}
	\caption{(a) -- Comparison of the effective rank of the MER array and the average effective rank of a random array, (b) --
	sensitivity ratio of the MER array to the average sensitivity of a random array.}
	\label{fig:6}
\end{figure}

The effective rank of MER array and the average effective rank of a random array are plotted in Fig. \ref{fig:6} along with the sensitivity ratios obtained at several frequencies.
It can be seen that when an array of only $2$ microphones is considered, the effective rank of the MER array is higher than the average effective rank of the random array by only $2$ dimensions; this difference increases to about $7$ dimensions when the number of microphones increases to $10$. The sensitivity ratio plot in Fig. \ref{fig:6}b shows that, as before, the MER array is more robust than the average random array, especially at lower frequencies. The improvement in array robustness increases to above $25$ dB for larger arrays. This is due to the larger increase in the gain in the effective rank for a larger number of microphones.

To conclude, the results above demonstrate that increasing the effective rank increases the array robustness, while the amount of improvement depends on the frequency and array size.

\subsection{DOA estimation performance} 
A simulation was carried out in order to evaluate the effect of the effective rank on the performance of the MUSIC DOA estimation algorithm. For this purpose, the estimation error was evaluated as a function of the four following parameters:
\begin{enumerate}
	\item frequency - $\{100,\,200,\,400,\,800,\,1600,\,3200\}$ Hz
	\item number of microphones - $\{2,\,3,\,4,\,5,\,7,\,10\}$
	\item $\mathrm{SNR}$ - $\{-20,\,-10,\,0,\,10,\,20,\,40\}$ dB
	\item array type: (i) MER, (ii) MER-$1$ dimension, (ii) MER-$5$ dimensions, (iv) random array
\end{enumerate}

In order to calculate the MUSIC variance for each of the $6\times 6\times 6\times 4=864$ above configurations, the array measurement was simulated as follows:
\begin{equation}
	\bb{p}=\bb{H}(\Omega_j)\bb{s}+\bb{n},
	\label{eq:54}
\end{equation}
where $\bb{H}(\Omega_j)\in\mathbb{C}^{L\times 1}$ describes the steering vector of the selected array (array type and number of microphones)  to a single arriving direction $\Omega_j$ at the selected frequency. The measurement noise $\bb{n}\in\mathbb{C}^{L}$ was obtained using the MATLAB \texttt{wgn()} function and thus was expected to be distributed jointly Gaussian with covariance $\sigma\bb{I}$. The SNR was controlled by setting the signal amplitude $\bb{s}$ to unity and the noise power $\sigma$ to the reciprocal of the desired SNR.

For each of the $864$ configurations the simulation followed several steps:
\begin{itemize}
	\item Select an arrival direction (out of $240$ nearly uniformly distributed directions) for the simulation of the
	measurement vector in \eqref{eq:54}.
	\item Estimate the measurement covariance matrix according to $\frac{1}{N}\sum\limits_{n=1}^N{\bb{p}\bb{p}^H}$ using $N=30$
	snapshots (differing by $\bb{n}$). Recall that the number of microphones satisfies $L<30$ in all configurations that were considered.
	\item Estimate the arrival direction using the MUSIC DOA estimation algorithm \cite{Stoica1989} , which can be one of the $240$
	directions for which the GHRTFs were obtained.
	\item Calculate the squared angle between the direction used in the simulation of $\bb{p}$ and the estimated arrival direction.
	\item Repeat the above steps $30$ times and average the resulting error.
	\item Repeat the above steps and average the resulting error over all of the $240$ directions.
	\item Repeat the above steps and average the resulting error over all of the $100$ realizations (i.e. different optimization 
	solutions) of the same array configuration (obtained as described in the previous section).
\end{itemize}

The quantity that was estimated in each configuration is the standard deviation (STD) of the angular error $\sqrt{\mathrm{E}[|\delta|^2]}$, where $\delta$ is the angle between the true direction and estimated arrival direction. Note that the possible values of $\delta$ are distributed in the range of $0^{\circ}-180^{\circ}$. In the case of selection by chance, i.e. when little information is available from the measurements, the distribution follows $\sin\delta$ as suggested by the surface differential $\sin\delta d\delta$. This implies that the probability density function of the absolute error is $f_{\delta}(\delta')=\frac{1}{2}\sin\delta'$ and the STD of $\delta$ is $\sim 98^{\circ}$. Thus, the values estimated here will be in the range of $0^{\circ}-98^{\circ}$ only, with $98^{\circ}$ being the selection by chance error.

\begin{figure}[ht] 
	\centering
	\includegraphics[width=\imgwidth\columnwidth]{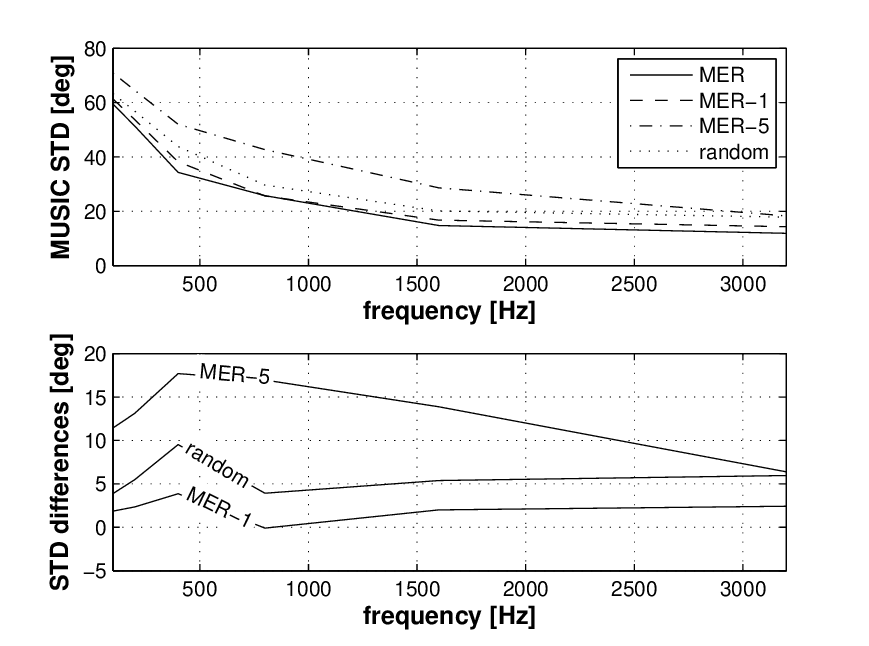} 
	\caption{MUSIC STDs and STD differences as a function of frequency and array	type; $L=3$, $\mathrm{SNR}=10$ dB.}
	\label{fig:7}
\end{figure}

Fig. \ref{fig:7} shows the STDs obtained in the simulation for different array types as a function of frequency.
It can be seen that the MER array has the lowest estimation error. Moreover, decreasing the effective rank generally results in larger STDs for all frequencies. In addition, the STD and the STD differences increase towards lower frequencies. This general behavior breaks at very low frequency, where the STD is very large and the differences become affected by the upper limit of $98^{\circ}$.

\begin{figure}[ht] 
	\centering
	\includegraphics[width=\imgwidth\columnwidth]{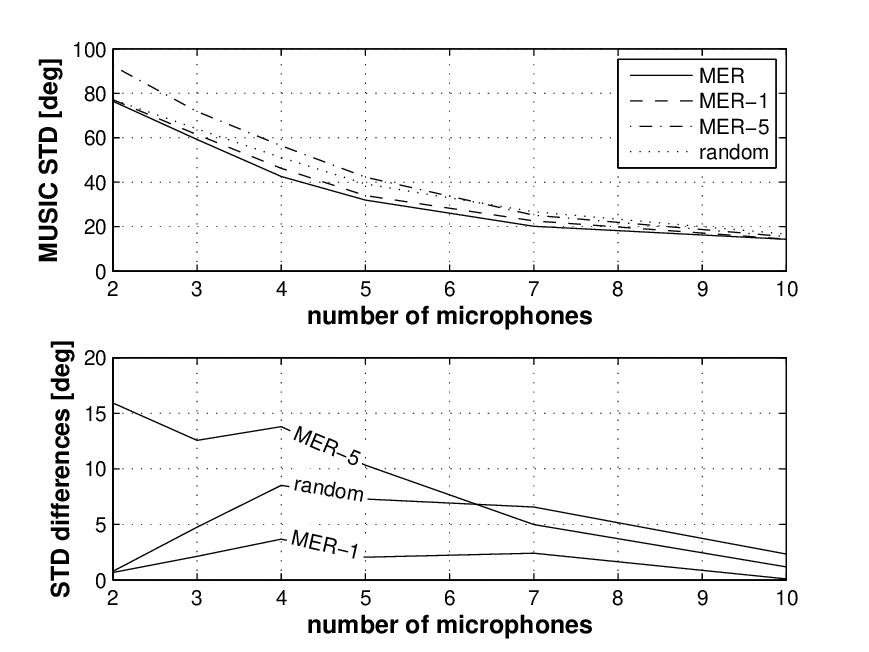} 
	\caption{MUSIC STD and STD differences as a function of array	size $L$, at the frequency of $400$ Hz and $\mathrm{SNR}=0$ dB.}
	\label{fig:8}
\end{figure}

Fig. \ref{fig:8} shows the MUSIC STDs and the STD differences as a function of array size.
Again, the best performance is obtained for the MER array and the performance degrades more in arrays with lower effective rank. The STDs and the STD differences generally increase for smaller arrays, except for $L=2,3$, where the behavior is probably affected by the upper error limit.

\begin{figure}[ht] 
	\centering
	\includegraphics[width=\imgwidth\columnwidth]{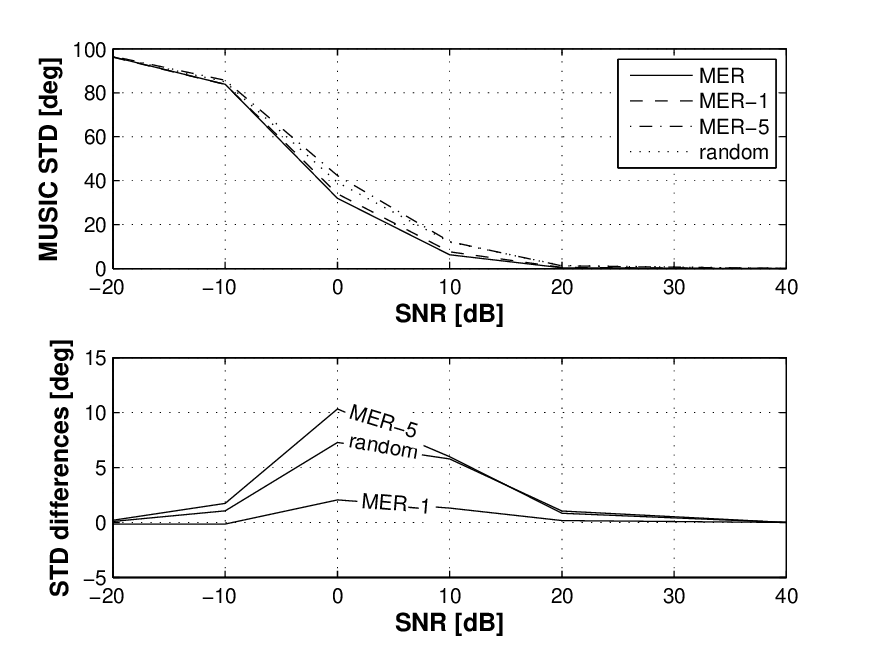} 
	\caption{MUSIC STD and STD differences as a function $\mathrm{SNR}$, at the frequency of $400$ Hz and for $L=5$.}
	\label{fig:9}
\end{figure}

Fig. \ref{fig:9} shows the MUSIC STDs and STD differences as a function of $\mathrm{SNR}$.
It can be seen that, once again, the MER array performs better than the others.
In addition, except for the cases of $\mathrm{SNR}=-10,-20$ dB, where the behavior is believed to be affected by the upper error limit, the STD differences increase for lower $\mathrm{SNR}$. This implies that the effect of the effective rank improvement is more salient at lower $\mathrm{SNR}$, which is in complete agreement with the results of the theoretical analysis (see Eq. \eqref{eq:213} and the discussion therein). 

In general, the MUSIC DOA estimation performance was demonstrated to improve when the effective rank of the array increased. It is emphasized that, ignoring the effect of the upper limit of $98^{\circ}$ on the estimation error, the improvement in performance is more significant for smaller arrays, lower frequencies and lower $\mathrm{SNR}$, where the available information is more limited.

	\section{Conclusion} 
A measure of microphone distribution quality for a microphone array for humanoid robot audition has been proposed. The measure was analyzed theoretically, showing that the positioning of the microphones according to this measure can generally improve the robustness of beamforming algorithms and reduce the MUSIC DOA estimation variance. The GA optimization approach was utilized in order to obtain examples of optimal array design. The examples were based on the GHRTF database, constructed for the surface of a human-like dummy head. Analyses of the examples are consistent with theoretical findings, demonstrating the advantages of array design using the proposed theoretical framework.

\section{Acknowledgments} 
The authors would like to thank Brian F. G. Katz from LIMSI-CNRS, France, for kindly providing the model of the head geometry that was used in this study, and for his valuable advice that helped to develop and validate the generalized HRTF database.

The research leading to these results has received funding from the European Union's Seventh Framework Programme (FP7/2007-2013) under grant agreement no. 609465.

	\bibliographystyle{IEEEtran}
	\bibliography{IEEEabrv,taslp1}
\begin{IEEEbiography} 
[{\includegraphics[width=1in,height=1.25in,clip,keepaspectratio]{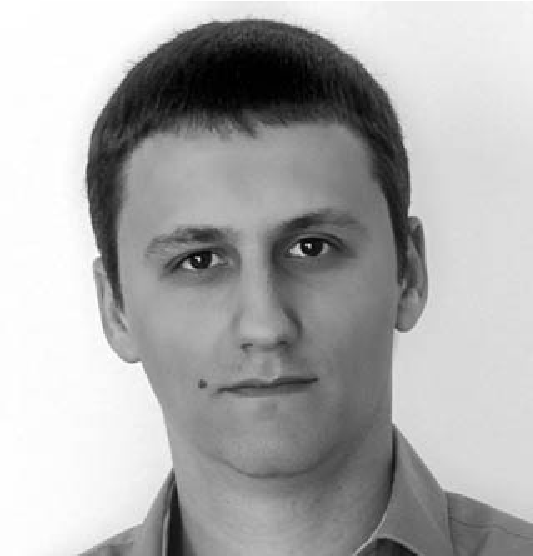}}]{Vladimir Tourbabin}
(S'12) received the B.Sc. degree (summa cum laude) in materials science and engineering and the M.Sc. degree (cum laude) in electrical and computer engineering from Ben-Gurion University of the Negev, Israel, in 2005 and 2011, respectively. He is currently working towards the Ph.D. degree in electrical and computer engineering at Ben-Gurion University.

His current research focuses on audition of humanoid robots.

Mr. Tourbabin is a recipient of the Negev Faran Fellowship.
\end{IEEEbiography}

\begin{IEEEbiography} 
[{\includegraphics[width=1in,height=1.25in,clip,keepaspectratio]{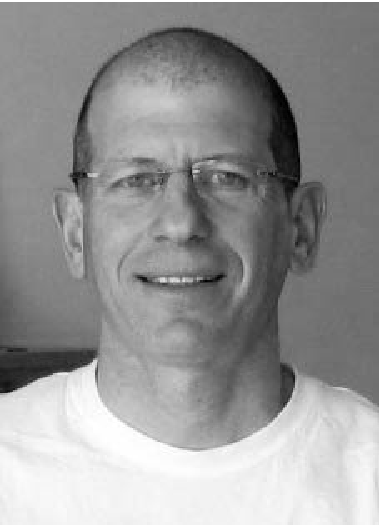}}]{Boaz Rafaely}
Boaz Rafaely (SM’01) received the B.Sc. degree (cum laude) in electrical engineering from Ben-Gurion University, Beer-Sheva, Israel, in 1986; the M.Sc. degree in biomedical engineering from Tel-Aviv University, Israel, in 1994; and the Ph.D. degree from the Institute of Sound and Vibration Research (ISVR), Southampton University, U.K., in 1997. 
At the ISVR, he was appointed Lecturer in 1997 and Senior Lecturer in 2001, working on active control of sound and acoustic signal processing. In 2002, he spent six months as a Visiting Scientist at the Sensory Communication Group, Research Laboratory of Electronics, Massachusetts Institute of Technology (MIT), Cambridge, investigating speech enhancement for hearing aids. He then joined the Department of Electrical and Computer Engineering at Ben-Gurion University as a Senior Lecturer in 2003, and appointed Associate Professor in 2010, and Professor in 2013. 

He is currently heading the acoustics laboratory, investigating sound fields by microphone and loudspeaker arrays. Since 2010, he is serving as an associate editor for IEEE Transactions on Audio, Speech and Language Processing, and since 2013 as a member of the IEEE Audio and Acoustic Signal Processing Technical Committee.

Prof. Rafaely was awarded the British Council’s Clore Foundation Scholarship.
\end{IEEEbiography}

\end{document}